\documentclass[letterpaper, 10 pt, conference]{ieeeconf}  

\IEEEoverridecommandlockouts                              

\usepackage{graphicx}
\usepackage{latexsym}
\usepackage{amsmath}
\usepackage{amssymb}
\usepackage{epsfig}
\usepackage{cite}
\usepackage{algorithmic}
\usepackage{overpic}
\usepackage{multirow}
\usepackage[table]{xcolor}
\usepackage{graphicx}
\usepackage{colortbl,array}
\usepackage{multirow,bigdelim}
\usepackage{graphicx}
\usepackage{subfigure}
\usepackage{booktabs}
\usepackage{booktabs,amsmath}
\usepackage{fancyhdr, graphicx, amsmath, amssymb}
\usepackage[linesnumbered,ruled]{algorithm2e}
\usepackage{arydshln}
\usepackage[english]{babel}
\usepackage{xcolor}

\SetCommentSty{mycommfont}

\newtheorem{definition}{Definition}

\newtheorem{lem}{Lemma}
\newtheorem{rem}{Remark}
\newtheorem{thm}{Theorem}

\newtheorem{asm}{Assumption}

\title{\LARGE \bf
Novel Defense Strategy Against Zero-Dynamics Attack in Multi-Agent Systems}

\author{Yanbing~Mao, Emrah~Akyol,
        and Ziang~Zhang
\thanks{Y.~Mao, E.~Akyol and Z.~Zhang are with the Department of Electrical and Computer Engineering, Binghamton University--SUNY, Binghamton, NY,
13902 USA {\tt\small {\{ymao3,eakyol,zhangzia\}}@binghamton.edu.} }
}

\begin{document}

\maketitle
\thispagestyle{empty}
\pagestyle{empty}

\begin{abstract}
This paper considers the defense strategy of strategic topology switching for the second-order multi-agent system under zero-dynamics attack (ZDA) whose attack-starting time is allowed to be not the initial time. We first study the detectability of ZDA, which is a sufficient and necessary condition of switching topologies in detecting the attack. Based on the detectability, a Luenberger observer under switching topology is proposed to detect the stealthy attack. The primary advantages of the attack-detection algorithm are twofold: (i) in detecting ZDA, the algorithm allows the defender (system operator) to have no knowledge of the attack-starting time and the misbehaving agents (i.e., agents under attack); (ii) in tracking system in the absence of attacks, Luenberger observer has no constraint on the magnitudes of observer gains and the number of observed outputs, i.e., only one agent's observed output is sufficient. Simulations are provided to verify the effectiveness of the strategic topology-switching algorithm.
\end{abstract}

\section{INTRODUCTION}
Coordination of multi-agent systems with the first-order dynamics is a well-studied theoretical problem with many practical applications. However, current and emerging systems,  such as vehicle~\cite{ren2007distributed}, spacecraft~\cite{abdessameud2009attitude}, robot~\cite{chung2009cooperative} and
electrical power networks~\cite{johnson2014synchronization}, rely on the second-order dynamics. A second-order multi-agent system consists of a population of $n$ agents whose dynamics are governed by the
following equations:
\begin{subequations}\label{eq:MA_sys}
\begin{align}
{\dot x_i}\left( t \right) &= {v_i}\left( t \right),\label{eq:oox1}\\
{{\dot v}_i}\left( t \right) &=  u_{i}(t),\hspace{0.5cm}i = 1, \ldots ,n\label{eq:oox2}
\end{align}\label{eq:ooiio}\end{subequations}
where $x_{i}(t) \in \mathbb{R}$ is the position, $v_{i}(t) \in \mathbb{R}$ is the velocity, and $u_{i}(t) \in \mathbb{R}$ is the local control input. Recently, significant efforts have been devoted to the coordination control of the second-order multi-agent system~(\ref{eq:ooiio}), see e.g., flocking~\cite{olfati2006flocking},  velocity synchronization and regulation of relative distances~\cite{tanner2007flocking}, decentralized formation control of spacecrafts~\cite{ren2004decentralized}, and distributed continuous-time optimization~\cite{rahili2017distributed}, etc. The numerous real networked systems that can be represented by~(\ref{eq:ooiio}) and the broad applications of its coordination control is the main motivation of this paper considering the model~(\ref{eq:ooiio}). For coordination control, this paper considers more representative average consensus.

Security concerns regarding networked cyber-physical systems pose an existential threat to their wide-deployment, see e.g., Stuxnet malware attack and Maroochy Shire Council Sewage control incident~\cite{A08}. The  ``networked" aspect  exacerbates the difficulty of detecting and preventing aforementioned attacks since centralized measurement (sensing) and control are not feasible for such large-scale systems~\cite{F13}, and hence requires the development of decentralized approaches, which are inherently prone to attacks. While developing defense strategies for ZDA have recently gained interest~\cite{n11,n12,F13,weerakkody2017robust,chen2017protecting}, the space of solutions is yet to be thoroughly explored. The most prominent features of prior work are that they constrain the connectivity of network topology and the number of the misbehaving agents (referred to agents under attack)~\cite{n11,n12,F13,weerakkody2017robust} or require the knowledge of the number of misbehaving agents and the attack-starting time at the defender (system operator) side for attack detection~\cite{n11,n12,F13,T12,n15,chen2017protecting}. The main objective of this work is to remove such constraints and unrealistic assumptions by utilizing a new approach for attack detection: intentional topology switching.

Recent experiment of stealthy false-data injection attacks on networked control system~\cite{T12} showed the changes in the system dynamics could be used to detect stealthy attack. To have changes in the system dynamics to detect ZDA, Teixeira et al.~\cite{T12} proposed a  method of modifying output matrix through adding and removing observed measurements, or modifying input matrix through adding and removing actuators or perturbing the control input matrix. In more realistic situation where the attack can exactly infer system dynamics and attack-starting
is designed to be not the initial time and the defender (system operator) has no knowledge of the attack-starting time, to detect ZDA the system dynamics must have dynamic changes, i.e., some parameters of system dynamics changes infinitely over infinite time.

For the dynamical networks, recent studies have highlighted the important role played by the network topology, and in recent several years, actively/strategically topology switching has received significantly attention in the control theory, network science and graph theory literatures, see e.g., Ciftcioglu et al.~\cite{ciftcioglu2017topology} studied dynamic topology design in the adversary environment where the network designer continually and strategically change network topology to a denser state, while the adversary attempts to thwart the defense strategy simultaneously. Moreover, driven by recent developments in mobile computing, wireless communication, sensing, etc., it is more feasible to set communication topology as a control variable~\cite{hartenstein2008tutorial,mazumder2011wireless}. These motivate us to consider the method of topology switching such that the multi-agent system can have changes in its system dynamics to detect ZDA.

The strategy on switching times proposed in~\cite{cdccs18} will be the building block of our defense strategy, which answered the question: \emph{when the topology of network should switch such that the occurring dynamic changes in system dynamics do not undermine the agent's ability of reaching consensus in the absence of attacks}. Based on the work in~\cite{cdccs18}, this paper focuses on the strategy on switching topologies that addresses the problem of \emph{switching to what topologies to detect ZDA}.

The contribution of this paper is threefold, which can be summarized as follows.
\begin{itemize}
  \item The detectability of ZDA is obtained, which is a sufficient and necessary condition of switching topologies.
  \item A sufficient and necessary condition of switching topologies for Luenberger observer is derived, which has no constraint on the number of observed outputs in tracking real multi-agent system.
  \item Through employing Luenberger observer, a strategic topology-switching algorithm for attack detection is proposed. The advantages of the algorithm are: i) in detecting ZDA, it allows the defender (system operator) to have no knowledge of misbehaving agents and the attack-starting time; ii) in tracking real systems in the absence of attacks, it has no constraint on the magnitudes of observer gains and  the number of observed outputs.
\end{itemize}

\section{Preliminaries}
\subsection{Notation}
$\mathbb{R}^{n}$ and $\mathbb{R}^{m \times n}$ denote the set of $\emph{n}$-dimensional real vectors and the set of $m \times n$-dimensional real matrices, respectively. Let $\mathbb{C}$ denote the set of complex numbers. $\mathbb{N}$ represents the set of the natural numbers and $\mathbb{N}_{0}$ = $\mathbb{N}$ $\cup$ $\left\{ 0 \right\}$. Let $\mathbf{1}_{n \times n}$ and $\mathbf{0}_{n \times n}$ be the $n \times n$-dimensional identity matrix and zero matrix, respectively. $\mathbf{1}_{n} \in \mathbb{R}^{n}$ and $\mathbf{0}_{n} \in \mathbb{R}^{n}$ denote the vector with all ones and the vector with all zeros, respectively. For a matrix $M \in \mathbb{R} ^{n \times n}$, $\lambda_{i}\left( M \right)$ denotes its $i^{\emph{\emph{th}}}$ eigenvalue. The superscript `$\top$' stands for matrix transpose. $\ker \left( Q \right) \triangleq \left\{ {y: Qy = {\mathbf{0}_n}}, Q \in \mathbb{R}^{n \times n} \right\}$. Also, $\left| \cdot \right|$ denotes the cardinality of a set, or the modulus of a number. ${\mathbb{V}} \backslash \mathbb{K}$ describes the complement set of $\mathbb{K}$ with respect to $\mathbb{V}$. $\mathbb{Q}$ denotes the set of rational numbers.

The interaction among $n$ agents is modeled by an undirected graph $\mathrm{G} \triangleq (\mathbb{V}, \mathbb{E})$, where
$\mathbb{V}$ $\triangleq$ $\left\{ {1,2, \ldots, n} \right\}$ is the set of vertices that represents $n$ agents and $\mathbb{E} \subseteq \mathbb{V} \times \mathbb{V}$ is the set of edges of the graph $\mathrm{G}$. The weighted adjacency matrix $\mathcal{A} = \left[ {{a_{ij}}} \right]$ $\in \mathbb{R}^{n \times n}$ of the graph $\mathrm{G}$ is defined as $a_{ij} = a_{ji} > 0$ if $(i, j) \in \mathbb{E}$, and $a_{ij} = a_{ji} = 0$ otherwise. Assume that there are no self-loops, i.e., for any ${i} \in \mathbb{V}$, $a_{ii} = 0$.  The Laplacian matrix of graph $\mathrm{G}$ is defined as $\mathcal{L} \triangleq \left[ {{l_{ij}}} \right] \in {\mathbb{R}^{n \times n}}$, where ${l_{ii}} \triangleq \sum\limits_{j = 1}^n {{a_{ij}}}$, and ${l_{ij}} \triangleq - {a_{ij}}$ for $i \neq j$.

\subsection{Attack Model}
As a class of stealthy attacks, ZDA is hard to detect, identify, and then mitigate from a control theory perspective~\cite{MP15,hl18,MP18}. Before reviewing its attack policy, let us first consider the following system:
\begin{subequations}
\begin{align}
\dot z\left( t \right) &= Az\left( t \right),\\
y\left( t \right) &= Cz\left( t \right),
\end{align}\label{eq:sl2}\end{subequations}
where $z(t) \in \mathbb{R}^{\bar{n}}$ and $y(t) \in \mathbb{R}^{\bar{m}}$ denote system state and observed output, respectively; $A \in \mathbb{R}^{\bar{n} \times \bar{n}}$, $C \in \mathbb{R}^{\bar{m} \times \bar{n}}$. Its corresponding version under attack is described by
\begin{subequations}
\begin{align}
\dot{\tilde z}\left( t \right) &= A\tilde z\left( t \right) + Bg(t),\\
\tilde y\left( t \right) &= C\tilde z\left( t \right) + Dg(t),
\end{align}\label{eq:sl1}\end{subequations}
where ${g}(t) \in \mathbb{R}^{\bar{o}}$ is attack signal, $B \in \mathbb{R}^{\bar{n} \times \bar{o}}$ and $D \in \mathbb{R}^{\bar{m} \times \bar{o}}$.

The policy of ZDA with introduction of attack-starting time is presented in the following definition, which is different from the attack policies studied in~\cite{n11,n12,F13,T12,n15,chen2017protecting}, whose attack-starting times are the initial time.
\begin{definition}~\cite{mao2017strategic} The attack signal
\begin{align}
{g}(t) = \left\{ \begin{array}{l}
\hspace{-0.2cm}{g}{e^{\eta \left( {t - \rho } \right)}},t \in \left[ {\rho,\infty } \right)\\
\hspace{-0.2cm}\mathbf{0}_{\bar{o}},\hspace{0.86cm}t \in \left[ {0,\rho } \right)
\end{array} \right.\label{eq:asg}
\end{align}
in system~(\ref{eq:sl1}) is a \emph{zero-dynamics attack}, if $\mathbf{0}_{\bar{n}} \neq \tilde{z}(0) - z(0) \in \mathbb{R}^{\bar{n}}$, $\mathbf{0}_{\bar{o}} \neq g(\rho) \in \mathbb{R}^{\bar{o}}$, $\rho \geq 0$ and $\eta \in \mathbb{C}$ satisfy
\begin{subequations}
\begin{align}
\!\!\tilde{z}\left( 0 \right) - z\left( 0 \right) &\!\in\! \ker \left( {{\mathcal{O}}} \right) \hspace{0.1cm} \emph{\emph{if}} \hspace{0.1cm}  \rho > 0\label{eq:aczeo}\\
\!\!\left[\!
    \begin{array}{c}
        {{e^{A\rho }}\left( {\tilde{z}\left( 0 \right) - z\left( 0 \right)} \right)}\\ \hdashline[2pt/2pt]
        { - g\left( \rho  \right)}
    \end{array}
\!\right] &\!\in\! \ker \left(\left[\!
    \begin{array}{c;{2pt/2pt}c}
        {\eta \mathbf{1}_{\bar{n} \times \bar{n}} - A} & B\\ \hdashline[2pt/2pt]
        -C & D
    \end{array}
\!\right]\right),\label{eq:czeo}
\end{align}\label{eq:ssczeo}
\end{subequations}where \begin{align}
{\mathcal{O}} \triangleq \left[
    \begin{array}{c;{1pt/1pt}c;{1pt/1pt}c;{1pt/1pt}c}
        {C^\top} & {{\left( {CA} \right)}^\top} & \ldots & {{\left( {C{A^{\bar{n} - 1}}} \right)}^\top}
    \end{array}
\right]^\top.\label{eq:1aczeo}
\end{align}
Moreover, the states and observed outputs of systems~(\ref{eq:sl1}) and~(\ref{eq:sl2}) satisfy
\begin{align}
y\left( t \right) &= \tilde y\left( t \right),\hspace{0.1cm} \emph{\emph{for}} \hspace{0.1cm} \emph{\emph{all}} \hspace{0.1cm} t \ge 0 \label{eq:rs1}\\
\tilde{z}\left( t \right) &= \left\{ \begin{array}{l}
\hspace{-0.2cm}{e^{At}}\tilde{z}\left( 0 \right), \hspace{2.98cm}t \in \left[ {0,\rho } \right]\\
\hspace{-0.2cm}z\left( t \right) + \left( {\tilde{z}\left( \rho  \right) - z\left( \rho  \right)} \right){e^{\eta \left( {t - \rho } \right)}}, t \in \left( {\rho,\infty } \right).
\end{array} \right.\label{eq:rs2}
\end{align}
\label{thm:zods}
\end{definition}

\begin{rem}
Compared with other stealthy attacks, see e.g., replay attack \cite{mo2009secure} and covert attack \cite{de2017covert}, ZDA has more explicit attack objectives. The state solution~(\ref{eq:rs2}) shows that through choosing the parameter $\eta$ and also the attack-starting time $\rho$, the attacker can achieve various objectives, see e.g.,
 \begin{itemize}
\item $\rho = \infty$: altering the steady-state value while not affecting system stability;
 \item $\rho < \infty$, ${\mathop{\rm Re}\nolimits} \left( \eta  \right)$ $>$ $0$: making system unstable;
 \item $\rho$ $<$ $\infty$, ${\mathop{\rm Re}\nolimits} \left( \eta  \right)$ $=$ $0$, ${\mathop{\rm Im}\nolimits} \left( \eta  \right)$ $\neq$ $0$: causing oscillatory behavior.
 \end{itemize}
The output~(\ref{eq:rs1}) indicates the undetectable/stealthy property.
\end{rem}

\subsection{Control Objective}
We recall the definition of second-order consensus to review the control objective.
\begin{definition}
\cite{yu2010some} The second-order consensus in the multi-agent system~(\ref{eq:ooiio}) is achieved if for any initial condition:  \begin{subequations}
\begin{align}
\mathop {\lim }\limits_{t \to \infty } | {{x_i}\left( t \right) - {x_j}\left( t \right)} | &= 0,\label{eq:oox1}\\
\mathop {\lim }\limits_{t \to \infty } | {{v_i}\left( t \right) - {v_j}\left( t \right)} | &= 0, \forall i,j = 1,
\ldots, n.\label{eq:oox2}
\end{align}\label{eq:defc}\end{subequations}
\end{definition}
We recall our proposed simplified control protocol that works for consensus in the environment of dynamic topology.
\begin{align}
u_i (t)  =  \sum\limits_{j \in \mathbb{V}} {a_{ij}^{\sigma \left( t \right)}\left( {{x_j}\left( t \right) - {x_i}\left( t \right)} \right)}, i \in \mathbb{V} \label{eq:lci}
\end{align}
where $\sigma (t):[t_{0},\infty ) \to \mathfrak{S} \triangleq \{1, \ldots, \mathrm{s}\}$, is the switching signal of the interaction topology of the communication network; $a^{\sigma(t)}_{ij}$ is the entry of the weighted adjacency matrix that describes the activated topology of communication graph.

\subsection{Strategy on Switching Times}
The following time-dependent topology switching is the building block of our defense strategy.
\begin{lem}~\cite{cdccs18}
Consider the second-order multi-agent system~(\ref{eq:ooiio}) with control protocol~(\ref{eq:lci}). For the given topology set $\mathfrak{S}$ that satisfies
\begin{align}
&\forall r \in \mathfrak{S}: \sqrt {\frac{{{\lambda _i}\left( {{{\cal L}_r}} \right)}}{{{\lambda _j}\left( {{{\cal L}_r}} \right)}}} \in \mathbb{Q}, \hspace{0.1cm} \emph{\emph{for}} \hspace{0.1cm} \forall i,j = 2, \ldots, \left| \mathbb{V} \right|\label{eq:bss1}
\end{align}
and the scalars $1 > \beta > 0$, $\alpha > 0$ and $\kappa \in \mathbb{N}$, if the dwell times $\tau_{r}$, $r \in \mathfrak{S}$, satisfy
\begin{align}
\tau_{r} = {{\widehat \tau_{\max} }} + \mathrm{m}\frac{{T}_{r}}{2}, \mathrm{m} \in \mathbb{N} \label{eq:addt}
\end{align}
where
$0$ $<$ ${{\widehat \tau_{\max} }}$ $<$ $\frac{{ - \ln \beta }}{\alpha }$, $0 < {{\widehat \tau_{\max} }} + {\frac{\mathrm{m}{T}_{r}}{2}} - \left( {{\beta ^{ - \frac{1}{\kappa}}} - 1} \right)\frac{\kappa}{{\alpha  - \xi }}$,
$\xi < \alpha$, $\xi = \mathop {\max }\limits_{r \in \mathfrak{S},i = 1, \ldots ,n} \left\{ {1 - {\lambda_i}\left( {{{\cal L}_r}} \right)}, {-1 + {\lambda_i}\left( {{{\cal L}_r}} \right)} \right\}$ and ${T}_{r}$ $=$ $\emph{\emph{lcm}}\left ( \frac{2\pi }{\sqrt{{\lambda_i}(\mathcal{L}_r})}; i = 2, ..., n \right )$,
then the asymptotic second-order consensus~(\ref{eq:defc}) is achieved.
\label{thm:lmr}
\end{lem}

\begin{rem}
For the control protocol, with undirected communication topology, that replies on both relative positions and velocities measurements, its obvious advantage is that there is no constraint on the magnitudes of coupling weights \cite{ren2007distributed}. The switching scheme proposed in Lemma \ref{thm:lmr} almost maintains this advantage since it only requires the rations of non-zero Laplacian eigenvalues to be rational numbers.
\end{rem}

\section{System Description}
For simplicity, we let the increasingly ordered set $\mathbb{M} \triangleq \{1, 2, \ldots\}$ $\subseteq \mathbb{V}$ denote the set of observed outputs. We usually refer to an agent under attack as a \emph{misbehaving agent}~\cite{n12}. We let $\mathbb{K}$ $\subseteq$ $\mathbb{V}$ denote the set of misbehaving agents.

Under time-dependent topology switching, the multi-agent system in \eqref{eq:ooiio} with the control input \eqref{eq:lci}  in the presence of attack and the observed outputs of positions can be written as
\begin{subequations}
\begin{align}
\!\!\!\!{{\dot{\tilde{x}}}_i}\!\left( t \right) &= {{\tilde v}_i}\!\left( t \right) \label{eq:oox1}\\
\!\!\!\!{{\dot{\tilde v}}_i}\!\left( t \right) &= \sum\limits_{i \in \mathbb{V}} \!{a_{ij}^{\sigma \left( t \right)}}\!\!\left( {{{\tilde x}_j}\!\left( t \right) \!-\! {{\tilde x}_i}\!\left( t \right)} \right) + \!\left\{ \begin{array}{l}
\hspace{-0.2cm}{\tilde{g}}_i\!\left( t \right)\!, i \!\in\! \mathbb{K}\\
\hspace{-0.2cm}0, \hspace{0.55cm} i \!\in\! {\mathbb{V}} \backslash \mathbb{K}
\end{array} \right.\label{eq:oox2}\\
\!\!\!\!{{\tilde{y}}_i}\!\left( t \right) &= \tilde x_{i}(t). i \!\in\! \mathbb{M}\label{eq:oox3}
\end{align}\label{eq:oofn}\end{subequations}

The system in (\ref{eq:oofn}) can be equivalently rewritten in the form of a switched system under attack:
\begin{subequations}
\begin{align}
&\dot{{{\tilde{z}}}}\left( t \right) = A_{{\sigma}(t)}{{\tilde{z}}}\left( t \right) + \tilde{g}\left( t\right)  \label{eq:s1a}\\
&\tilde{y}\left( t \right) = C\tilde{z}(t),\label{eq:s1b}
\end{align}\label{eq:s1}\end{subequations}
where
\begin{subequations}
\begin{align}
\!\!\! \!\!\tilde{z}\left( t \right) &\!\triangleq\! \left[
    \begin{array}{c;{1pt/1pt}c;{1pt/1pt}c;{1pt/1pt}c;{1pt/1pt}c;{1pt/1pt}c}
        \!\!\!{{\tilde{x}_1}\left( t \right)} \!\!&\! {\ldots} \!\!&\! {\tilde{x}_{\left| \mathbb{V} \right|}}\left( t \right) \!\!&\! {\tilde{v}_1}\left( t \right) \!\!&\! {\ldots} \!\!&\! {\tilde{v}_{\left| \mathbb{V} \right|}}\left( t \right)
    \end{array}\!\!\!
\right]^\top\!,  \label{eq:ssd1}\\
 \!\!\! \!\!A_{\sigma(t)} &\!\triangleq\! \left[
    \begin{array}{c;{2pt/2pt}c}
        \mathbf{0}_{\left| \mathbb{V} \right| \times \left| \mathbb{V} \right|} & \mathbf{1}_{\left| \mathbb{V} \right| \times \left| \mathbb{V} \right|}\\ \hdashline[2pt/2pt]
        -\mathcal{L}_{\sigma(t)} & \mathbf{0}_{\left| \mathbb{V} \right| \times \left| \mathbb{V} \right|}
    \end{array}
\right],\label{eq:nm0} \\
 \!\!\!\!\!{C} &\!\triangleq\! \left[
    \begin{array}{c;{1pt/1pt}c}
   \!\!\!\emph{\emph{diag}}\!\left\{ 1, \ldots , 1 \right\} \!\!\!&\!\! \mathbf{0}_{\left| \mathbb{M} \right| \times \left( {\left| \mathbb{V} \right| - \left| \mathbb{M} \right|} \right)}\!\!\!\!
    \end{array}\right],\label{eq:u1b}\\
        \!\!\!\!\!\tilde{g}(t) &\!\triangleq\! \left[\!\!
    \begin{array}{c;{1pt/1pt}c}
        {\bf{0}}_{\left| \mathbb{V} \right|}^\top & \bar{g}^{\top}(t)
    \end{array}
\!\!\right]^\top,\label{eq:nm2}\\
\!\!\!\!\![{\bar{g}}\left( t \right)]_{i} &\!\triangleq\! \left\{ \begin{array}{l}
\hspace{-0.2cm}{\tilde{g}_i}\left( t \right),i \in \mathbb{K}\\
\hspace{-0.2cm}0, \hspace{0.62cm}i \in {\mathbb{V}} \backslash \mathbb{K}.
\end{array} \right.\label{eq:u1}
\end{align}
\end{subequations}In addition, we consider the system \eqref{eq:s1} in the absence of attacks, which is given by
\begin{subequations}
\begin{align}
\dot{{{{z}}}}\left( t \right) &= A_{{{\sigma}}(t)}{{{z}}}\left( t \right),\label{eq:s21}\\
y\left( t \right) &= C{z}(t).\label{eq:s22}
\end{align}\label{eq:s2}\end{subequations}

To end this section, we make the following assumptions on the attacker and the defender (system operator).
\begin{asm}
The attacker
\begin{enumerate}
  \item knows the currently activated topology and its dwell time at switching time;
  \item has the memory of the past switching sequences.
\end{enumerate}\label{thm:att}
\end{asm}
\begin{asm}
The defender
\begin{enumerate}
  \item designs the switching sequences including switching times and topologies;
  \item  has no knowledge of the attack-starting time;
  \item  has no knowledge of the number of misbehaving agents.
\end{enumerate} \label{thm:auatt}
\end{asm}

\section{Detectability of Zero-Dynamics Attack}
This section focuses on the detectability of ZDA, which will answer the question: \emph{what topologies of multi-agent system~(\ref{eq:oofn}) to strategically switch to such that the attack policy~(\ref{eq:ssczeo}) is not feasible?}

To better illustrate the strategy on switching topologies, we introduce the definition of components in a graph.
\begin{definition}[Components of Graph~\cite{NW10}] The \emph{components} of a graph $\mathcal{G}$ are its maximal connected subgraphs. A component is said to be \emph{trivial} if it has no edges; otherwise, it is a \emph{nontrivial component}.
\end{definition}

\begin{figure}
\centering
\includegraphics[scale=0.6]{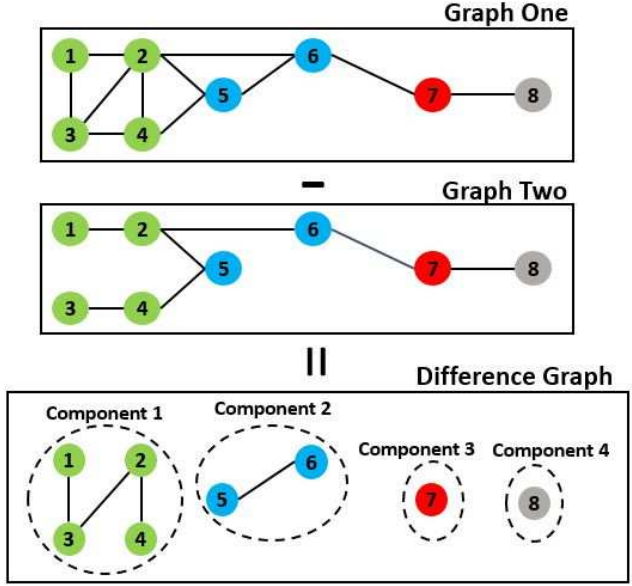}
\caption{Components of difference graph (the weights of communications links are uniformly set as ones).}
\label{fig:cpp}
\end{figure}

\begin{rem}
As illustrated by Difference Graph in Fig.~\ref{fig:cpp}, different components do have any common vertex, otherwise they are in the same component formed via that common vertex.
\end{rem}

\begin{definition}
The difference graph $\mathrm{G}^{rs}_{\emph{\emph{diff}}} = \left( {\mathbb{V}_{\emph{\emph{diff}}}^{rs},\mathbb{E}_{\emph{\emph{diff}}}^{rs}} \right)$ of two graphs $\mathrm{G}_{r}$ and $\mathrm{G}_{s}$ is generated as
\begin{align}
\mathbb{V}_{\emph{\emph{diff}}}^{rs} &= {\mathbb{V}_r} \cup {\mathbb{V}_s} \nonumber\\
\left( {i,j} \right) &\in \mathbb{E}_{\emph{\emph{diff}}}^{rs}, \hspace{0.1cm}\emph{\emph{if}} \hspace{0.1cm} a_{ij}^r - a_{ij}^s \ne 0 \nonumber
\end{align}
where $\mathbb{V}_r$ and $a^{r}_{ij}$ are the set of agents and the entry of weighted adjacency matrix of the graph $\mathrm{G}_{r}$, respectively.
\end{definition}

We define the union difference graph for switching difference graphs:
\begin{align}
{\mathrm{G}_{\emph{\emph{diff}}}} &\triangleq \left( {\bigcup\limits_{r,s \in \mathfrak{S}} {\mathbb{V}_{\emph{\emph{diff}}}^{rs}} ,\bigcup\limits_{r,s \in \mathfrak{S}} {\mathbb{E}_{\emph{\emph{diff}}}^{rs}} } \right).\label{eq:udg}
\end{align}

We use $\mathbb{C}_{i}({\mathrm{G}_{\emph{\emph{diff}}}})$ to denote the set of agents in the $i^{\emph{\emph{th}}}$ component of union difference graph ${\mathrm{G}_{\emph{\emph{diff}}}}$. Obviously, $\mathbb{V}$ $=$ $\mathbb{C}_{1}$ $\bigcup$ $\mathbb{C}_{2}$ $\bigcup$ $\ldots$ $\bigcup$ $\mathbb{C}_{\mathrm{d}}$, and $\mathbb{C}_{\mathrm{p}}$ $\bigcap$ $\mathbb{C}_{\mathrm{q}}$ $=$ $\emptyset$ if $\mathrm{p}$ $\neq$ $\mathrm{q}$, where $\mathrm d$ is the number of the component of difference graph ${\mathrm{G}_{\emph{\emph{diff}}}}$.

The detectability of ZDA under strategy on switching topologies are presented in the following theorem.
\begin{thm}~\cite{mao2017strategic} Consider the multi-agent system~(\ref{eq:s1}). Under time-dependent topology switching, the ZDA can be detected by defender without knowledge of misbehaving agents and the attack-starting time, if and only if at least one agent in each component of union difference graph has observed output, i.e.,
  \begin{align}
  {\mathbb{C}_i}({\mathrm{G}_{\emph{\emph{diff}}}}) \cap \mathbb{M} \ne \emptyset ,\forall i = 1, \ldots \mathrm{d}.\label{eq:cccc}
  \end{align}
\label{thm:sms}
\end{thm}

\section{Attack Detection Algorithm}
Based on the obtained detectability of ZDA, this section focus on its detection algorithm.
\subsection{Luenberger Observer under Switching Topology}
We now present a Luenberger observer for the system~(\ref{eq:oofn}):
\begin{subequations}
\begin{align}
\dot{\mathbf{x}}_i\left( t \right) &= \mathbf{v}_i\left( t \right)\label{eq:fl1}\\
\dot{\mathbf{v}}_i\left( t \right) &= \sum\limits_{i \in \mathbb{V}} \!{a_{ij}^{\sigma \left( t \right)}}\!\!\left( {{\mathbf{x}_j}\!\left( t \right) - {\mathbf{x}_i}\!\left( t \right)} \right) \nonumber\\
&\hspace{2.55cm} -  \left\{ \begin{array}{l}
\hspace{-0.2cm}\psi_{i}{r_{i}(t)} \!+\! \theta_{i}{\dot{r}_{i}(t)}, i \!\in\! \mathbb{M}\\
\hspace{-0.2cm}0, \hspace{1.7cm} i \!\in\! {\mathbb{V}} \backslash \mathbb{M}
\end{array} \right. \\
r_i(t) &= \mathbf{x}_i(t) - {{\breve{y}_{i}}}\left( t \right), i \!\in\! \mathbb{M} \label{eq:fl3}
\end{align}\label{eq:fl}\end{subequations}
where ${\tilde{y}}_{i}(t)$ is the observed output in system~(\ref{eq:oofn}), $r_{i}\left( t  \right)$ is the attack-detection signal, $\psi_{i}$ and $\theta_{i}$ are the observer gains designed by the defender (system operator).

The strategy~(\ref{eq:s2}) in Theorem~\ref{thm:sms} implies that if the union difference graph is connected, i.e., the union difference graph has only one component, using only one agent's observed output is sufficient to detect ZDA. The following result regarding the stability of system~(\ref{eq:fl}) will answer the question: \emph{whether only one agent's observed output is sufficient for the observer to asymptotically track the system~(\ref{eq:s1}) in the absent of attack? }
\begin{thm}~\cite{mao2017strategic}
Consider the following matrix:
\begin{align}
\mathcal{A}_{s} \triangleq \left[
    \begin{array}{c;{1.0pt/1.0pt}c}
        {\mathbf{0}_{n\times n}} & {\mathbf{1}_{n\times n}}  \\ \hdashline[1pt/1pt]
        { - \mathcal{L}_{s} - \Phi} & {-\Theta}
    \end{array}
\right],\label{eq:dm}
\end{align}
where
\begin{subequations}
\begin{align}
\Phi  &\triangleq \emph{\emph{diag}}\{ {\psi_{1}, \ldots, \psi_{\left| \mathbb{M} \right|}, 0, \ldots, 0} \},\label{eq:flvb1}\\
\Theta &\triangleq \emph{\emph{diag}}\{{\theta_{1}, \ldots, \theta_{\left| \mathbb{M} \right|}, 0, \ldots, 0} \}.\label{eq:flvb2}
\end{align}\label{eq:flp}\end{subequations}
Given that $\mathcal{L}_{s}$ is the Laplacian matrix of a connected graph and the gain matrices $\Phi$ and $\Theta$ satisfy
\begin{subequations}
\begin{align}
\mathbf{0}_{n \times n} \neq \Phi \geq 0,\label{eq:aaa}\\
\mathbf{0}_{n \times n} \neq \Theta \geq 0,\label{eq:bbb}
\end{align}
\end{subequations}
then $\mathcal{A}_{s}$ is Hurwitz for any $\left| \mathbb{M} \right| \ge 1$, if and only if $\mathcal{L}_{s}$ has distinct eigenvalues. \label{thm:my0bd}
\end{thm}

\subsection{Strategic Switching Topology For Detection}
The strategic topology-switching algorithm is described by Algorithm~1.
\begin{algorithm}
  \caption{Strategic Topology-Switching Algorithm}
  \KwIn{Initial index $k$ = 0, initial time $t_{k} = 0$, an ordered topology set $\mathfrak{S}$ that satisfies
  \begin{align}
  \exists s \in \mathfrak{S}: \mathcal{L}_{s} \hspace{0.1cm} \emph{\emph{has}} \hspace{0.1cm} \emph{\emph{distinct}} \hspace{0.1cm} \emph{\emph{eigenvalues}},\label{eq:bss2}
  \end{align}
  dwell times $\tau_{s}$, $s \in \mathfrak{S}$, generated by~(\ref{eq:addt}) that satisfy
  \begin{align}
  \sum\limits_{s \in \mathfrak{S}} {{\nu _s}{\mu _P}\left( {{{\cal A}_s}} \right)}  < 0,{\nu _s} = \frac{{{\tau _s}}}{{\sum\limits_{r \in \mathfrak{S}} {{\tau _r}} }}.\label{eq:bss2}
  \end{align}}
  Run the multi-agent system~(\ref{eq:s1}) and the observer~(\ref{eq:fl})\;
  Switch topology of system~(\ref{eq:oofn}) and its observer~(\ref{eq:fl}) at time $t_{k} + \tau_{\tilde{\sigma}(t_{k})}$: $\tilde \sigma ({t_k}) \leftarrow \mathfrak{S}\left( \!\!\!\!{\mod \!\!\!\left( {k + 1,\left| \mathfrak{S} \right|} \right) + 1} \right)$\;
  Update topology-switching time: $t_{k} \leftarrow t_{k} + \tau_{\tilde{\sigma}(t_{k})}$\;
  Update index: $k \leftarrow k+1$\;
  Go to Step 2.
\end{algorithm}

\begin{thm}~\cite{mao2017strategic}
Consider the multi-agent system~(\ref{eq:s1}) and the observer~(\ref{eq:fl}), where the observer gain matrices $\Phi$ and $\Theta$ satisfy~(\ref{eq:aaa}) and~(\ref{eq:bbb}), and the topology-switching signal $\tilde{\sigma}(t_{k})$ of the observer~(\ref{eq:fl}) and the system~(\ref{eq:s1}) are generated by Algorithm~1.
\begin{description}
  \item[i)] Without knowledge of the misbehaving agents and the attack-starting time, the observer~(\ref{eq:fl}) is able to detect ZDA in system~(\ref{eq:s1}), i.e., $r(t) \equiv \mathbf{0}_{\left| \mathbb{M} \right|}$ does not holds, if and only if the set of observed outputs and switching topologies satisfy~(\ref{eq:cccc}).
  \item[ii)] In the absence of attacks, without constraint on the magnitudes of observer gains, the observer~(\ref{eq:fl}) asymptotically track the real system~(\ref{eq:oofn}).
  \item[iii)]In the absence of attacks, the agents in system~(\ref{eq:s1}) achieve the asymptotic second-order consensus.
\end{description}
\label{thm:dfd}
\end{thm}

\section{Simulation}
\begin{figure}[http]
\centering
\includegraphics[scale=0.40]{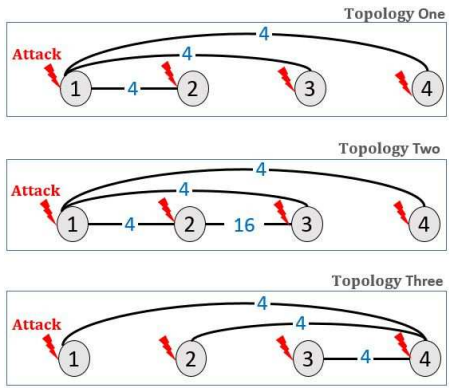}
\caption{Working situation.}
\label{fig:dp}
\end{figure}

We consider a system with $n = 4$ agents. The initial position and velocity conditions are chosen randomly as ${x}(0) = {v}(0) = {\left[ {1,2,3,4} \right]^ \top }$. The considered network topologies with their coupling weights are given in Fig.~\ref{fig:dp}, which shows that all of the four agents are under attack, i.e., $\mathbb{K} = \left\{ {{1},{2},{3},{4}} \right\}$. We let agent 1 provide observed output, i.e.,
 $\mathbb{M} = \left\{ {{1}} \right\}$. We set the observer gains significantly small as $\Phi = \Theta = \emph{\emph{diag}}\{10^{-6},0,0,0\}.$ We note that in the working situation illustrated by Fig.~\ref{fig:dp}, the existing results~\cite{n11,n12,F13,weerakkody2017robust,chen2017protecting} for the multi-agent systems under fixed topology fail to detect ZDA. This is mainly due to the misbehaving-agents set $|\mathbb{K}|= 4$; the connectivities of Topologies One, Two and Three are as the same as $1$; and the output set $|\mathbb{M}| = 1$. All these violate the conditions on the connectivity of the communication network, the size of the misbehaving-agent set,  and the size of the output set.

First, we consider the topology set $\mathfrak{S} = \{1, 2\}$, and set the topology switching sequence as $1 \to 2 \to 1 \to 2 \to  \ldots$, periodically. We verify from Fig.~\ref{fig:dp} that the topology set $\mathfrak{S} = \{1, 2\}$ satisfies~(\ref{eq:bss1}) and~(\ref{eq:bss2}). By Lemma~\ref{thm:lmr}, we select the dwell times $\tau_{1} =  \tau_{2} = \frac{T_{1}}{2} + 0.2 = \frac{T_{2}}{2} + 0.2$ = $\frac{\pi}{2} + 0.2$. It can be verified from Topologies One and Two in Fig.~\ref{fig:dp} that their generated difference graph is disconnected. Thus, the set $\mathfrak{S} = \{1, 2\}$ does not satisfy~(\ref{eq:cccc}) in Theorem~\ref{thm:sms}. Therefore, the attacker can easily design a ZDA such that Algorithm~1 fails to detect it. Following the policy~(\ref{eq:ssczeo}), one of the stealthy attack strategies is designed as:
\begin{itemize}
  \item $\eta = 0.0161$;
  \item modify the data of initial condition sent to observer~(\ref{eq:fl}) as $\widehat{x}\left( 0 \right)$ = ${\left[ {1,1,3,5} \right]^\top}$ and $\widehat{v}\left( 0 \right)$ = ${\left[ {1,1,4,4} \right]^\top}$;
  \item choose attack-starting time $\rho = 1097.4$;
  \item introduce ZDA signal to system at $\rho$: \\
  $g\left( t \right) \!=\! {10^{ - 3}}{\left[ {0,{\rm{7}}{\rm{.3}}{e^{\eta (t - \rho )}},{\rm{7}}{\rm{.3}}{e^{\eta (t - \rho )}}, - {\rm{14}}{\rm{.6}}{e^{\eta (t - \rho )}}} \right]^\top}$.
\end{itemize}

\begin{figure}[http]
\centering{
\begin{minipage}[b]{0.55\textwidth}
\includegraphics[width=0.8\textwidth]{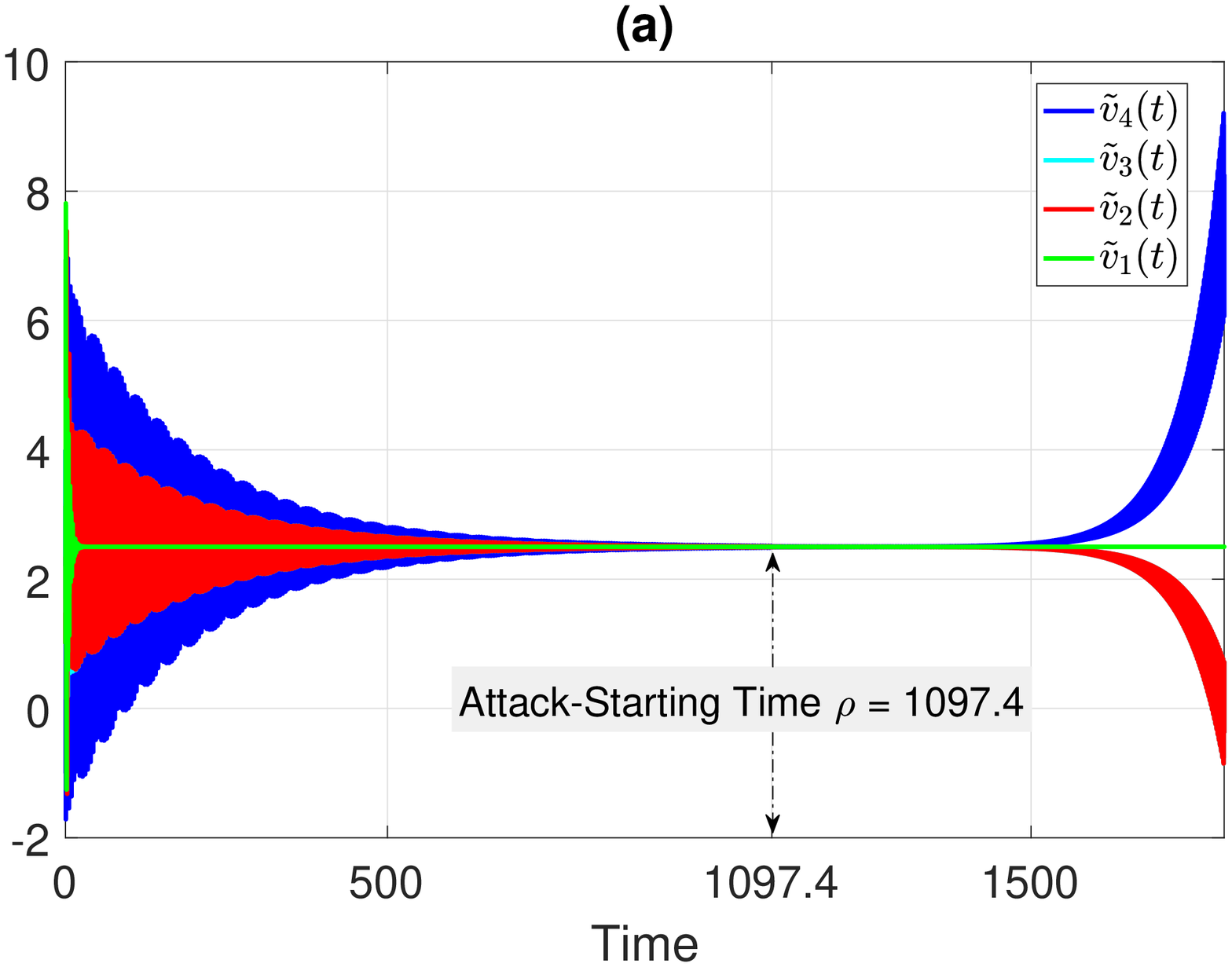} \\
\includegraphics[width=0.8\textwidth]{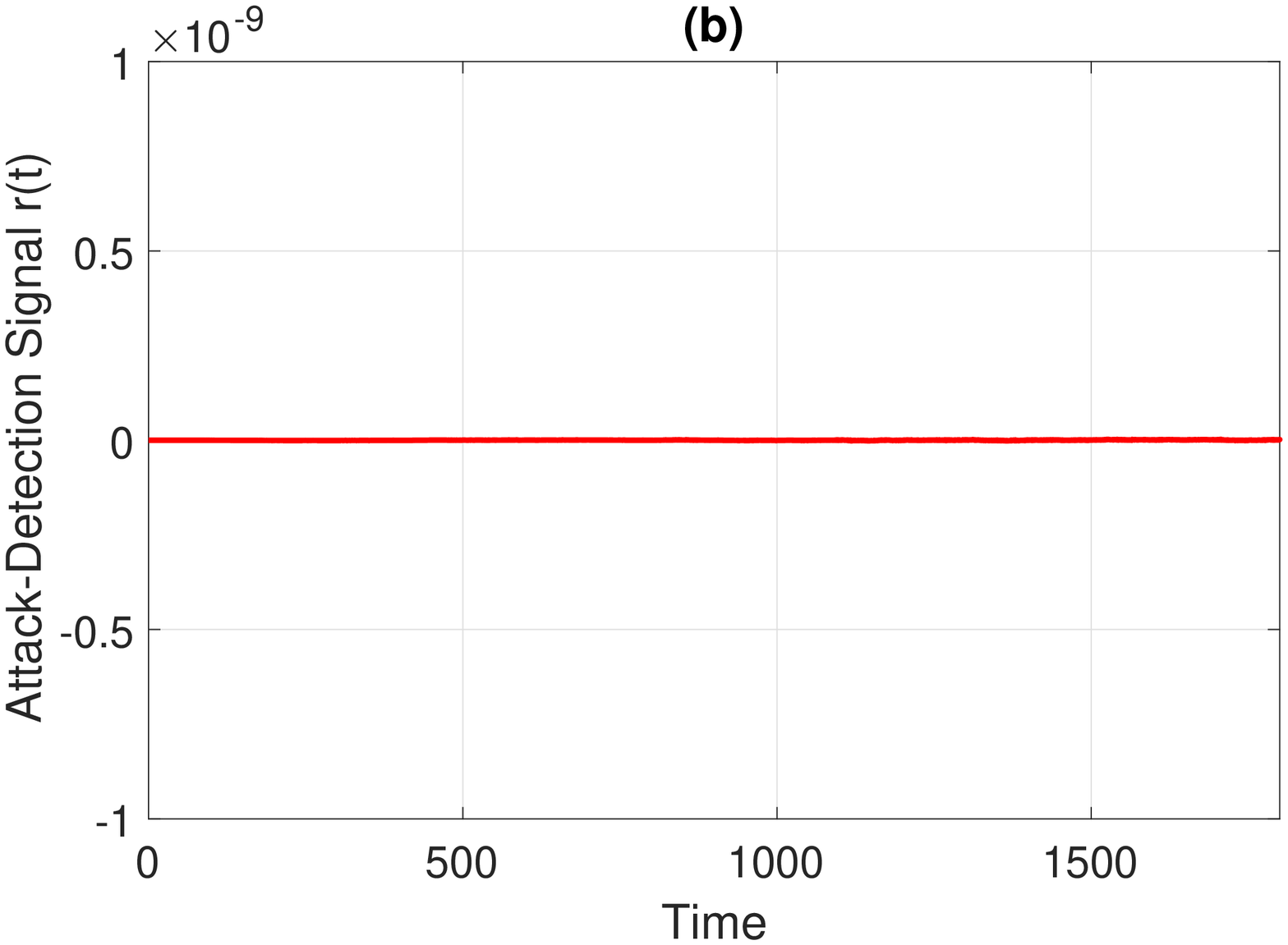}
\end{minipage}}
\caption{States $\tilde{v}(t)$: multi-agent system under attack is unstable; attack-detection signal $r(t)$: the attacker keep stealthy over time.}
\label{fig:trp}
\end{figure}

The trajectories of detection signal $r(t)$ designed in~(\ref{eq:fl}), and the velocities are shown in Fig.~\ref{fig:trp}, which illustrates that the attacker's goal of making the system unstable without being detected is achieved under the topology set $\mathfrak{S} = \{1, 2\}$.
\begin{figure}[http]
\centering{
\includegraphics[scale=0.35]{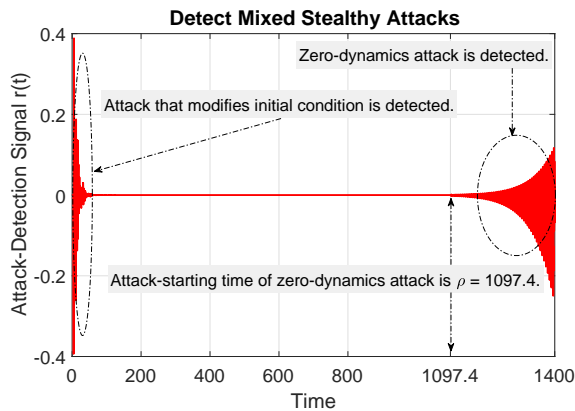}}
\caption{Attack-detection signal $r(t)$: using only one agent's observed outputs, the designed ZDA is detected.}
\label{fig:rev}
\end{figure}

To detect the designed stealthy attack, we now incorporate Topology Three in Fig.~\ref{fig:dp} into topology set, i.e., $\mathfrak{S} = \{1, 2,3\}$. We let the topology switching sequence to be $1 \to 2 \to 3 \to 1 \to 2 \to 3 \to \ldots$, periodically. We verify that the topology set $\mathfrak{S} = \{1, 2,3\}$ satisfies~(\ref{eq:bss1}) and~(\ref{eq:bss2}). Using Lemma~\ref{thm:lmr}, the dwell times are selected as $\tau_{1} =\tau_{2} = \tau_{3} = \frac{\pi}{2} + 0.2$. It can be verified from Fig.~\ref{fig:dp} that the difference graph generated by Topologies One and Three, or  Topologies Two and Three is connected. Thus, by property i) in Theorem~\ref{thm:dfd}, we conclude that using only one agent's observed output, the strategic topology-switching algorithm--Algorithm~1, is able to detect the designed ZDA under the topology set $\mathfrak{S} = \{1, 2,3\}$. The trajectory of the attack-detection signal $r(t)$ in Fig.~\ref{fig:rev} illustrates that with all of the agents being misbehaving, using only one agent's observed outputs, Algorithm~1 succeeds in detecting the mixed stealthy attacks.

\section{Conclusion}
In this paper, we propose a strategy on switching topologies that addresses the problem: \emph{what topology to switch to, such that the ZDA can be detected}. Based on the strategy, a defense strategy is derived. The theoretical results obtained in this paper imply a rather interesting result:
for the dwell time of switching topologies, there exist a tradeoff between the switching cost and the duration of attacks going undetected, and the convergence speed to consensus.
Analyzing the tradeoff problems in the lights of game theory and multi-objective optimization constitutes a part of our future research.

\section{Acknowledgment}
This work has been supported in part by Binghamton University--SUNY, Center for Collective Dynamics of Complex Systems ORC grant.

\bibliographystyle{IEEEtran}
\bibliography{refII}
\end{document}